\documentclass[journal]{IEEEtran}
\usepackage{amsmath,amssymb,amsfonts}
\usepackage{cite}
\usepackage{graphicx}
\usepackage{bm}
\usepackage{textcomp}
\usepackage{booktabs} 
\usepackage{ragged2e}

\newcommand {\Define} {\stackrel {\Delta} {=}  }

\begin{document}

\title{Low-complexity Equalization of Zak-OTFS Via Neumann Series}

\author{Vineetha Yogesh\IEEEauthorrefmark{1}, Saif Khan Mohammed\IEEEauthorrefmark{2}\IEEEauthorrefmark{3},  Sandesh Rao Mattu\IEEEauthorrefmark{3}, Ronny Hadani\IEEEauthorrefmark{4}\IEEEauthorrefmark{3} and Robert Calderbank\IEEEauthorrefmark{1}
\IEEEauthorblockA{\IEEEauthorrefmark{1}Department of Electrical and Computer Engineering, Duke University, USA}\\
\IEEEauthorblockA{\IEEEauthorrefmark{2}Department of Electrical Engineering, Indian Institute of Technology Delhi, New Delhi, India}\\
\IEEEauthorblockA{\IEEEauthorrefmark{3}Cohere Technologies Inc., San Jose, CA, USA}
\IEEEauthorblockA{\IEEEauthorrefmark{4}Department of Mathematics, University of Texas at Austin, USA}\\
\thanks{S. K. Mohammed is currently with Cohere Technologies Inc., San Jose, CA, USA, on extra-ordinary leave from Department of Electrical Engineering, Indian Institute of Technology Delhi, New Delhi, India}}

\maketitle

\begin{abstract}
We describe a general method for selecting an orthonormal basis of carrier waveforms that aligns the basis with delay / Doppler characteristics of a wireless channel. We show that our method enables low-complexity equalization for two channels of practical interest. The first is satellite communication and the second is communication from a ground station to an unmanned aerial vehicle (UAV). After Doppler compensation both scenarios are characterized by a first line of sight (LOS) path with zero delay and zero Doppler shift, and a second weaker path. We show that our method is robust to fractional delay and Doppler shifts.  

We consider orthonormal bases of carrier waveforms that are obtained from the pulsone basis of Zak-OTFS carrier waveforms by applying a generalization of the discrete affine Fourier transform (GDAFT). This family of waveforms includes AFDM and other modulations proposed for 6G. What distinguishes these bases is that the carrier waveforms are common eigenvectors of some maximal commutative subgroup S of a Heisenberg-Weyl group of discrete delay and Doppler shifts. We describe how to choose S to mitigate the damaging effects of interference between carriers. We represent the wireless channel by delay-Doppler taps and observe that a channel tap located within S multiplies every waveform by a complex phase. If all channel taps are located within S, then the channel multiplies every waveform by a complex phase, and a single tap equalizer supports reliable communication. This is the case for a linear time-invariant (LTI) channel, where S is the group of discrete time shifts and the carrier waveforms are discrete tones (OFDM). In general, we choose the subgroup S to maximize the diagonal component of the channel energy. In this case we describe how to use a Neumann Series to design an approximate equalizer, and we show that for our two channels of practical interest, it improves upon the single-tap equalizer.
\end{abstract}


\section{Introduction}
We consider a discrete-time model for signals that are limited to time $T$ and bandwidth $B$. We partition the duration $T$ into $N$ equal subintervals and the bandwidth $B$ into $M$ equal subintervals, so that the number of degrees of freedom is the time-bandwidth product BT=MN. Section~\ref{sec2} introduces Zak-OTFS modulation and it describes how channels and pulse shaping filters act by twisted convolution on Zak-OTFS carriers in the delay-Doppler (DD) domain. It then derives the input-output (I/O) relation in discrete-time and in matrix-vector form. 

Section~\ref{sec3} describes how orthonormal bases associated with Zadoff Chu (ZC) sequences are obtained from the pulsone basis of Zak-OTFS carrier waveforms by applying a generalization of the discrete affine Fourier transform (GDAFT). It then describes how the maximal commutative subgroups corresponding to these bases have the geometric structure of lines modulo MN, with the ZC root determining the slope of the line. When all the channel taps fall on a line modulo $MN$, the channel acts diagonally on the corresponding carrier waveforms, and Section~\ref{sec3} describes the single tap equalizer that results. 

When the taps do not fall on a line, the action of the wireless channel on the corresponding carrier waveforms will not be diagonal, but the action may be diagonally dominant. In this case, Section~\ref{sec4} describes how to construct an approximation to the Zero-Forcing (ZF) equalizer which we call the Neumann Equalizer (NE), because it is based on convergence of Neumann series. 

Section~\ref{sec5} provides simulation results for a 2-path channel where, after preconditioning, the first line-of-sight path has zero delay and zero Doppler shift, and the second weaker path has positive delay and non-zero Doppler shift. We present bit error rate (BER) results comparing performance of the single-tap, Neumann, Zero-Forcing and MMSE equalizers. We consider fractional delay and Doppler shifts and we also illustrate how BER performance changes as a function of the mismatch between the tap location and the integer grid.  

\section{System Model and the Zak-OTFS I/O relation}
\label{sec2}
The DD spreading function of the doubly-selective channel between the transmitter and the receiver is given by \cite{Bello, Hlawatsch2011}
\begin{eqnarray}
\label{eqn1}
    h(\tau, \nu)  = \sum\limits_{i=1}^{P} h_i \delta(\tau - \tau_i) \delta(\nu - \nu_i),
\end{eqnarray}where $P$ is the number of channel paths and $h_i, \tau_i, \nu_i$ denote the complex gain, delay and Doppler shift of the $i$-th channel path, $i=1,2,\cdots, P$.
The transmitted time-domain (TD) signal is denoted by $x(t)$ and the received TD signal is given by \cite{Bello}
\begin{eqnarray}
\label{eqn2}
    r(t)  =  \iint h(\tau, \nu) x(t - \tau) \, e^{j 2 \pi \nu (t - \tau)} \, d\tau \, d\nu \, + \, n(t),
\end{eqnarray}where $n(t)$ is AWGN. 
Next, we briefly discuss Zak-OTFS modulation and its I/O relation in the DD domain and the time-domain (TD) (for further details on Zak-OTFS and foundations please refer to \cite{zakotfs1, zakotfs2, otfsbook}).

In Zak-OTFS modulation, the carrier waveforms are quasi-periodic pulses in the DD domain, with period $\tau_p$ and $\nu_p \Define 1/\tau_p$ along the delay and Doppler axis respectively.
Let $B$ and $T$ denote the bandwidth and time of the transmitted signal $x(t)$. Let $M \Define B \tau_p$ and $N \Define T \nu_p$ be integers.
The information is first embedded in the discrete DD domain signal $x_{dd}[k,l], k,l \in {\mathbb Z}$ which is quasi-periodic with period $M$ and $N$ along the discrete delay and Doppler axis respectively, i.e. for all $n,m \in {\mathbb Z}$
\begin{eqnarray}
\label{eqn3}
    x_{dd}[k + nM, l + mN] & = & e^{j 2 \pi \frac{ n l}{N}} \, x_{dd}[k,l].
\end{eqnarray}From (\ref{eqn3}) it is clear that the $MN$ information symbols, $x_{dd}[k_0,l_0], k_0=0,1,\cdots, M-1, l_0=0,1,\cdots, N-1$ are repeated periodically along the delay and Doppler axis. $x_{dd}[k,l]$ is then lifted to the information lattice $\Lambda \Define \{ (k \tau_p/M, l \nu_p/N) \, | \, k,l \in {\mathbb Z} \}$ resulting in the continuous DD domain signal
\begin{eqnarray}
\label{eqn4}
    \Tilde{x}_{dd}(\tau, \nu)  =  \sum\limits_{k,l \in {\mathbb Z}} x_{dd}[k,l] \, \delta(\tau - k\tau_p/M) \, \delta(\nu - l \nu_p/N).
\end{eqnarray}To limit the transmit signal $x(t)$ in time and bandwidth, $\Tilde{x}_{dd}(\tau, \nu)$ is filtered with the pulse shaping filter $w_{tx}(\tau, \nu)$ resulting in the filtered signal $x_{dd}(\tau,\nu)$ which is given by
\begin{eqnarray}
\label{eqn5}
    x_{dd}(\tau, \nu) & = & w_{tx}(\tau, \nu) *_{\sigma} \Tilde{x}_{dd}(\tau, \nu), \nonumber \\
    & & \hspace{-25mm} =  \iint w_{tx}(\tau', \nu') \Tilde{x}_{dd}(\tau - \tau', \nu - \nu') \, e^{j 2 \pi \nu' (\tau - \tau')} \, d\tau' \, d\nu',
\end{eqnarray}where $*_{\sigma}$ denotes the twisted convolution operation. The transmitted signal is the time-domain (TD) representation of $x_{dd}(\tau, \nu)$ which is given by its inverse Zak transform (${\mathcal Z}_t^{-1}$) \cite{Zak67, Janssen88, DerivOTFS, TDDDconversion}
\begin{eqnarray}
\label{eqn6}
    x(t) & = & {\mathcal Z}_t^{-1}(x_{dd}(\tau, \nu)) = \sqrt{\tau_p} \int\limits_{0}^{\nu_p} x_{dd}(t, \nu) \, d\nu.
\end{eqnarray}At the receiver, the DD representation of $r(t)$ is given by its Zak transform (${\mathcal Z}_t$) 
\begin{eqnarray}
\label{eqn7}
    r_{dd}(\tau, \nu)  =  {\mathcal Z}_t\left(r(t) \right) = \sqrt{\tau_p} \sum\limits_{n \in {\mathbb Z}} r(\tau + n \tau_p) \, e^{-j 2 \pi n \nu \tau_p}.
\end{eqnarray}The received DD signal is then match-filtered with the receive filter $w_{rx}(\tau, \nu) \Define w_{tx}^*(-\tau, -\nu)$ resulting in the received filtered DD signal
\begin{eqnarray}
\label{eqn8}
    y_{dd}(\tau, \nu) & = & w_{rx}(\tau, \nu) *_{\sigma} r_{dd}(\tau, \nu).
\end{eqnarray}Sampling on the information lattice $\Lambda$ gives the received discrete DD domain signal
\begin{eqnarray}
\label{eqn9}
    y_{dd}[k,l] & \Define & y_{dd}\left( \tau = \frac{k \tau_p}{M} \,,\, \nu = \frac{l \nu_p}{N} \right).
\end{eqnarray}From equations (\ref{eqn1}) through (\ref{eqn9}), the Zak-OTFS input-output (I/O) relation in the DD domain is given by
\begin{eqnarray}
\label{eqn10}
    y_{dd}[k,l] & = & h[k,l] *_{\sigma} x_{dd}[k,l] \, + \, n_{dd}[k,l],\nonumber \\
    h[k,l] & \Define & h_{dd}\left( \tau = \frac{k \tau_p}{M} \,,\, \nu = \frac{l \nu_p}{N} \right), \nonumber \\
    h_{dd}(\tau, \nu) & = &  w_{rx}(\tau, \nu) *_{\sigma} h(\tau, \nu) *_{\sigma} w_{tx}(\tau, \nu),
\end{eqnarray}where
\begin{eqnarray}
\label{eqn11}
    h[k,l] *_{\sigma} x_{dd}[k,l] & & \nonumber \\
    & & \hspace{-25mm} = \sum\limits_{k',l' \in {\mathbb Z}} h[k',l'] \, x_{dd}[k - k', l- l'] \, e^{j 2 \pi \frac{l'(k - k')}{MN}},
\end{eqnarray}is the discrete twisted convolution between $h[k,l]$ and $x_{dd}[k,l]$. Also, in (\ref{eqn10}), the AWGN DD samples $n_{dd}[k,l]$ are obtained by sampling the filtered AWGN $w_{rx}(\tau, \nu) *_{\sigma} n_{dd}(\tau,\nu)$ on the information lattice $\Lambda$ where $n_{dd}(\tau, \nu) \Define {\mathcal Z}_t\left( n(t) \right)$.

The discrete-time domain I/O relation is given by taking the Inverse Discrete Zak Transform (IDZT) of both sides of the discrete DD I/O relation in (\ref{eqn10}), which gives \cite{NishantDiscreteRadar, otfsbook}
\begin{eqnarray}
\label{eqn12}
    y[n] & \Define & \text{IDZT}(y_{dd}[k,l]) \, = \frac{1}{\sqrt{N}} \sum\limits_{l=0}^{N-1} y_{dd}[n,l], \nonumber \\
    & & \hspace{-5mm} = \sum\limits_{k,l \in {\mathbb Z}} h[k,l] \, x[n - k] \, e^{j 2 \pi \frac{l (n - k)}{MN}} \, + \, w[n], \nonumber \\
    x[n] & \Define & \text{IDZT}(x_{dd}[k,l]) \,,\,
    w[n] \, \Define \, \text{IDZT}(n_{dd}[k,l]).
\end{eqnarray}Note that $x[n], y[n], w[n]$ are all periodic discrete-time signals with period $MN$.
The average received signal to noise ratio (SNR) is given by
\begin{eqnarray}
\label{eqn13}
    \gamma  \Define  \frac{\sum\limits_{n=0}^{MN-1} {\mathbb E}\left[ {\Big \vert} \sum\limits_{k,l \in {\mathbb Z}} h[k,l] x[n - k] \, e^{j 2 \pi \frac{l (n - k)}{MN}}  {\Big \vert}^2 \right]}{\sum\limits_{n=0}^{MN-1} {\mathbb E}[\vert w[n] \vert^2 ]}.
\end{eqnarray}$x_{dd}[k,l]$ can be expressed in terms of the $MN$ information symbols as
\begin{eqnarray}
\label{eqn2974688}
    x_{dd}[k,l] & = & \sum\limits_{k_0=0}^{M-1} \sum\limits_{l_0 = 0}^{N-1} x_{dd}[k_0, l_0] \, \phi_{dd,k_0, l_0}[k,l],
\end{eqnarray}where $\phi_{k_0, l_0}[k,l]$ is the quasi-periodic DD domain signal which carries the information symbol $x_{dd}[k_0, l_0]$, $k_0=0,1,\cdots,M-1$, $l_0 = 0,1,\cdots, N-1$, and is given by
\begin{eqnarray}
    \phi_{dd,k_0, l_0}[k,l] & & \nonumber \\
    & & \hspace{-20mm} \Define \sum\limits_{n,m \in {\mathbb Z}} e^{j 2 \pi \frac{l_0 n}{N} } \, \delta[k - k_0 - nM] \, \delta[l - l_0 - mN].
\end{eqnarray}Taking IDZT of both sides of (\ref{eqn2974688}) gives us an expression of $x[n]$ in terms of the $MN$ information symbols $x_{dd}[k_0, l_0]$
\begin{eqnarray}
    x[n] & = & \sum\limits_{k_0 = 0}^{M-1}\sum\limits_{l_0 = 0}^{N-1} x_{dd}[k_0, l_0] \, \phi_{k_0, l_0}[n],
\end{eqnarray}where
\begin{eqnarray}
    \phi_{k_0, l_0}[n] & \Define & \text{IDZT}(\phi_{dd,k_0, l_0}[k,l]), \nonumber \\
    & & \hspace{-20mm} = \frac{1}{\sqrt{N}}\sum\limits_{i \in {\mathbb Z}} e^{j 2 \pi \frac{i l_0}{N}} \, \delta[n - k_0 - iM], \nonumber \\
    & & \hspace{-20mm} = \frac{1}{\sqrt{N}} \, e^{j 2 \pi \frac{l_0 (n - k_0)}{MN}} \sum\limits_{i \in {\mathbb Z}} \delta[n - k_0 - iM],
\end{eqnarray}is the discrete-time representation of the $(k_0, l_0)$-th Zak-OTFS carrier waveform which carries the $(k_0, l_0)$-th information symbol $x_{dd}[k_0, l_0]$. Note that the Zak-OTFS carrier $\phi_{k_0, l_0}[n]$ is essentially a pulse train modulated by a sinusoid and is therefore referred to as a pulsone.

The I/O relation in (\ref{eqn12}) can be expressed in matrix-vector form as follows. We organize $x[n], y[n], w[n], n=0,1,\cdots, MN-1$ into vectors ${\bf x}, {\bf y}, {\bf w}$ respectively which are related by
\begin{eqnarray}
\label{eqn14}
    {\bf y} & = & {\bf H} {\bf x} \, + \, {\bf w},
\end{eqnarray}where ${\bf H} \in {\mathbb C}^{MN \times MN}$ denotes the effective channel matrix, whose element in its $n$-th row and $m$-th column, $n,m = 0,1,\cdots, MN-1$ is given by
\begin{eqnarray}
\label{eqn15}
    H_{n,m} & = & \sum\limits_{q,l \in {\mathbb Z}} h[(n-m) + qMN, l] \, e^{j 2 \pi \frac{m l}{MN}}.
\end{eqnarray}A more useful decomposition of ${\bf H}$ is as follows. Let ${\mathcal S}$ denote the support set of $h[k,l]$, i.e., DD taps $(k,l)$ for which $\vert h[k,l] \vert^2$ is a significant fraction of the total channel power $\sum\limits_{k,l \in {\mathbb Z}} \vert h[k,l] \vert^2$.
\begin{eqnarray}
\label{eqn151}
    {\mathcal S} & \Define & \{ (k_p, l_p) \, | \, p = 1,2,\cdots, N_t \},
\end{eqnarray}where $N_t$ is the number of significant DD taps of $h[k,l]$. Then we can write $h[k,l]$ as
\begin{eqnarray}
    \label{eqn152}
    h[k,l] & = & \sum\limits_{p=1}^{N_t} h_p \, \delta[k - k_p] \, \delta[l - l_p].
\end{eqnarray}The $p$-th term corresponds to a Doppler shift by $l_p$ Doppler taps followed by a delay shift of $k_p$ delay taps. A delay by $k_p$ taps transforms the signal $x[n]$ to $x[n - k_p]$. The vector corresponding to $x[n - k_p]$ is $(x[MN - k_p] , x[MN - k_p + 1], \cdots, x[MN - 1 - k_p])^T$ is simply ${\bf P}^{k_p} \, {\bf x}$, where ${\bf P} \in {\mathbb C}^{MN \times MN}$ is the permutation matrix which transforms the vector ${\bf x} = (x[0], \cdots, x[MN-1])^T$ to $(x[1], x[2], \cdots, x[MN -1], x[0])^T$. Similarly, a Doppler shift by $l_p$ Doppler taps to the vector ${\bf x}$ results in the vector $(x[0] , x[1] e^{j 2 \pi \frac{l_p}{MN} }, \cdots, x[n] e^{j 2 \pi \frac{n l_p}{MN}} , \cdots, x[MN-1] e^{j 2 \pi \frac{(MN-1)l_p}{MN}})^T$ which is simply ${\bf M}^{l_p}$, where ${\bf M}$ is the $MN \times MN$ diagonal matrix whose $(n,n)$-th diagonal element is $e^{j 2 \pi \frac{n}{MN}}$, $n=0,1,\cdots,MN-1$. The $p$-th tap $(k_p, l_p)$ transforms the input ${\bf x}$ to $h_p {\bf P}^{k_p} {\bf M}^{l_p} {\bf x}$.
Hence an equivalent representation of ${\bf H}$ is given by
\begin{eqnarray}
  \label{eqn153}
  {\bf H} & = & \sum\limits_{p=1}^{N_t} h_p  {\bf P}^{k_p} {\bf M}^{l_p} .
\end{eqnarray}

\section{Single-tap equalization for channel taps on a line}
\label{sec3}
If all taps $(k_p, l_p) \in {\mathcal S}$ lie on a line ${\mathcal L}_u = \{ (k,l) \, \vert \, l = -u k (\text{mod} \, MN) \}$, for some $u$ relatively prime to $MN$ ($M$ and $N$ are odd primes) then it is possible to choose carrier waveforms which are eigen-vectors of the channel, i.e., the information symbols carried by these carrier waveforms do not interfere, thereby enabling separate equalization which has low complexity (i.e., single-tap equalization).

To be precise, consider the following Zadoff-Chu (ZC) basis of $MN$ discrete-time carrier waveforms (each capable of carrying one information symbol). ZC sequences are known for their desirable auto-correlation properties and are used in wireless standards like 3GPP LTE and 5G New Radio (NR) \cite{Andrews22}.
The $q$-th discrete-time $MN$-periodic waveform in this basis, $q=0,1,\cdots, MN-1$, is given by
\begin{eqnarray}
\label{eqn16}
    z_{u,q}[n] & \Define & \frac{1}{\sqrt{MN}} \, e^{-j \pi u \frac{(n - q)(n- q+1)}{MN}}, n \in {\mathbb Z}.
\end{eqnarray}One can check that shifting the waveform $z_{u,q}[n], n \in {\mathbb Z}$ by any point $(k_p, l_p) \in {\mathcal L}_u$ gives $z_{u,q}[n], n \in {\mathbb Z}$ itself. By direct calculation
\begin{eqnarray}
\label{eqn17}
    Z_{u,q}[n - k_p] \, e^{j 2 \pi \frac{l_p(n - k_p)}{MN}} & = & c_{k_p,u,q} \, Z_{u,q}[n], n \in {\mathbb Z}, \nonumber \\
    c_{k_p,u,q} & \Define & e^{j \pi \frac{u}{MN} (k_p^2 + k_p - 2qk_p)}.
\end{eqnarray}Let ${\bf z}_{u,q} = (z_{u,q}[0], \cdots, z_{u,q}[MN-1])^T$ be the vector notation for the waveform $z_{u,q}[n]$. Consider the matrix ${\bf Z}_u \in {\mathbb C}^{MN \times MN}$ whose $q$-th column is ${\bf z}_{u,q}$. From the properties of the ZC sequence, ${\bf Z}_u$ is unitary, i.e.
\begin{eqnarray}
\label{eqn18}
    {\bf Z}_u^H {\bf Z}_u & = & {\bf I}.
\end{eqnarray}We note that the basis of Zak-OTFS $MN$ carriers $\phi_{k_0, l_0}[n],$ $k_0 =0,1,\cdots, M-1$, $l_0=0,1,\cdots, N-1$ is also a orthonormal basis just like the above orthonormal basis of ZC waveforms for $MN$-periodic discrete-time signals. In fact we can obtain the ZC basis from the Zak-OTFS pulsone basis by applying the Generalized Discrete Affine Fourier Transform (GDAFT) which is a unitary transformation \cite{NishantSpread}.
Equation (\ref{eqn17}) is equivalent to
\begin{eqnarray}
    \label{eqn19}
    {\bf P}^{k_p} \, {\bf M}^{l_p} \, {\bf z}_{u,q} & = & c_{k_p,u,q} \, {\bf z}_{u,q},
\end{eqnarray}$q=0,1,\cdots, MN-1$.
Using (\ref{eqn19}) in (\ref{eqn153}) gives
\begin{eqnarray}
    \label{eqn20}
    {\bf H}  \, {\bf z}_{u,q} & = & \lambda_{u,q}\, {\bf z}_{u,q}, \,\,\,
    \lambda_{u,q} \, \Define \,   \sum\limits_{p=1}^{N_t} h_p \,  c_{k_p,u,q} ,
\end{eqnarray}i.e., the waveforms ${\bf z}_{u,q}, q=0,1,\cdots, (MN-1)$ are eigen-vectors of the effective channel matrix ${\bf H}$, when the effective channel taps $(k_p, l_p) \in {\mathcal L}_u$, $p=1,2,\cdots, N_t$. Let $s_q, q=0,1,\cdots, MN-1$ denote the $q$-th information symbol carried by the q-th waveform $z_q[n]$. The transmitted vector is therefore
\begin{eqnarray}
\label{eqn21}
    {\bf x} & = & {\bf Z}_u \, {\bf s},
\end{eqnarray}where ${\bf s} = (s_0, \cdots, s_{MN-1})^T$ is the vector of information symbols. Using this in (\ref{eqn14}) gives
\begin{eqnarray}
\label{eqn22}
    {\bf y} & = & {\bf H} \, {\bf Z}_u \, {\bf s} \, + \, {\bf w}.
\end{eqnarray}At the receiver, we compute
\begin{eqnarray}
\label{eqn23}
    {\bf r} & \Define & {\bf Z}_u^H \, {\bf y} \, = \, {\bf H}_{u} \, {\bf s} \, + \, {\bf w}_u, \nonumber \\
    {\bf H}_u & \Define & {\bf Z}_u^H \, {\bf H} \, {\bf Z}_u\,\,,\,\, {\bf w}_u \, \Define \, {\bf Z}_u^H \, {\bf w}.
\end{eqnarray}When the channel taps are on the line ${\mathcal L}_u$, from (\ref{eqn20}) we have
\begin{eqnarray}
\label{eqn24}
    {\bf H} \, {\bf Z}_u & = & {\bf Z}_u \, {\bf \Lambda}_u,
\end{eqnarray}where ${\bf \Lambda}_u$ is a $MN \times MN$ diagonal matrix whose $(q,q)$-th element is $\lambda_{u,q}$, $q=0,1,\cdots, (MN-1)$. Using (\ref{eqn24}) in (\ref{eqn23}) then gives
\begin{eqnarray}
\label{eqn25}
    {\bf H}_u & = & {\bf \Lambda}_u, \,\, \text{i.e.},  \,\,\, 
    {\bf r} \, = \, {\Lambda}_u \, {\bf s} \, + \, {\bf w}_u.
\end{eqnarray}In other words, the information symbols can be detected separately, i.e., single-tap equalization (STE) which has complexity only $O(MN)$. The output of the single-tap equalizer for the $q$-th information symbol is simply $r_q/\lambda_{u,q}$, where $r_q$ is the $q$-th element of ${\bf r}$. In a practical implementation, the mathematical operations of converting $r(t)$ to $y[n]$ can be carried out through continuous-time filtering and windowing (corresponding to the receiver matched-filter) followed by sampling at $B$ Hz and then periodizing the discrete-time sampled output with period $MN$, gives $y[n]$ (see \cite{Zak_OTFS_over_CP_OFDM} for details). The projection of the received ${\bf y}$ onto the ZC basis, i.e., ${\bf r} = {\bf Z}_u^H {\bf y}$ can be computed with complexity $O(MN \log (MN))$, and therefore the overall receiver complexity is $O(MN \log(MN))$.\footnote{\footnotesize{The $q$-th element of ${\bf r}$ in (\ref{eqn23}) is ${\bf z}_{u,q}^H {\bf y}$ which can be simplified and written as $\frac{e^{j 2 \pi u \frac{(q^2 - q)}{MN}}}{\sqrt{MN}}  \sum\limits_{n=0}^{MN-1} \hspace{-3mm} e^{-j 2 \pi n \frac{q u}{MN}} \, e^{j \pi u \frac{(n^2 + n)}{MN}} y[n]$, $q=0,1,\cdots, MN-1$. This can be computed by taking the product of $y[n]$ with $e^{j \pi u \frac{(n^2 + n)}{MN}}$ which has $O(MN)$ complexity, followed by the DFT of this product sequence which has $O(MN \log(MN))$ complexity. The DFT output is then permuted since $uq$ in the DFT sum exponent permutes $q=0,1,\cdots, MN-1$ to $uq = 0, u, 2u, ...., u(MN-1)$ since $u$ is relatively prime to $MN$. The permuted DFT output is then multiplied with $e^{j 2 \pi u \frac{(q^2 - q)}{MN}}$ which has $O(MN)$ complexity. }} 

\section{Proposed Neumann Equalization (NE)}
\label{sec4}
When the taps do not lie on a line, then ${\bf H}_u$ is non-diagonal and the information symbols interfere with each other. However, in many scenarios it may be possible to choose a $u$ such that ${\bf H}_u$ is diagonally dominant.
Let
\begin{eqnarray}
\label{eqn26}
    {\bf H}_u & = & {\bf D}_u \left({\bf I} - {\bf E}_u \right),
\end{eqnarray}where ${\bf D}_u$ is the diagonal matrix whose diagonal elements are same as that of ${\bf H}_u$. 
Strict row-wise diagonal dominance of ${\bf H}_u$ implies that the off-diagonal matrix ${\bf E}_u$ has spectral radius less than one \cite{Horn13}. We exploit this fact to propose a Neumann equalizer which approximately computes the Zero-Forcing (ZF) equalizer ${\bf H}_u^{-1}$.
\begin{eqnarray}
    \label{eqn27}
    {\bf H}_u^{-1} = \left( {\bf I} - {\bf E}_u \right)^{-1} {\bf D}_u^{-1}.
\end{eqnarray}When the spectral radius of ${\bf E}_u$ is less than one, the ``Neumann series" $S_n = \sum\limits_{i=0}^n {\bf E}_u^i$ converges to $({\bf I} - {\bf E}_u)^{-1}$ as $n \rightarrow \infty$ \cite{Horn13}.
\begin{eqnarray}
    \left( {\bf I} - {\bf E}_u \right)^{-1}  & = & \sum\limits_{i=0}^{\infty} {\bf E}_u^i.
\end{eqnarray}An approximation to the ZF equalizer output is therefore obtained by restricting the sum in the RHS above to the first $R$ terms, and is given by
\begin{eqnarray}
    \widehat{\bf s} & = & {\bf D}_u^{-1} \left( \sum\limits_{i=0}^{R-1} {\bf E}_u^i \right) \, {\bf r}.
\end{eqnarray}The complexity of this Neumann equalizer is $O(R M^2N^2)$ which is significantly less than the $O(M^3 N^3)$ complexity of the exact ZF equalizer ${\bf H}_u^{-1} \, {\bf r}$. Simulation results in Section \ref{sec5} reveal that for a $2$-tap channel, even when we have fractional taps which do not lie exactly on any line, there exists $u$ for which the performance of the Neumann equalizer is better than STE and is close to the performance of the ZF equalizer.

\section{Simulation results}
\label{sec5}
We consider a two-path channel where the first line-of-sight (LOS) path has zero delay and zero Doppler shift and the second weaker path has positive delay and non-zero Doppler shift. This models the channel for scenarios like UAV to satellite communication or communication between a ground station and a UAV, where the strong LOS path allows for accurate estimation and compensation of the Doppler shift on this path \cite{Itu24, tworay, Haas, TDDDconversion}. If the Doppler shifts of the LOS and the reflected path are $\nu_1$ and $\nu_2$ respectively, then after Doppler compensation, the effective Doppler shifts are $0$ and $(\nu_2 - \nu_1)$ respectively. The weaker path models the reflection of the signal from ground/sea in case of UAV-to-satellite communication, or reflection from a mountain/hill in case of UAV-to-ground communication. With $P=2$ paths and $\tau_1 = \nu_1 = 0$, the channel DD spreading function in (\ref{eqn1}) is

{\vspace{-4mm}
\small
\begin{eqnarray*}
    h(\tau, \nu)  =  \sqrt{\frac{K}{(K + 1)}}  \delta(\tau) \delta(\nu) + \sqrt{\frac{1}{(K + 1)}} \, h_2 \delta(\tau - \tau_2) \delta(\nu - \nu_2),
\end{eqnarray*}\normalsize}where $K$ is the Rician factor, and $h_2$ is a zero mean complex Gaussian circularly symmetric random variable with zero mean and unit variance.
We consider the Gauss-sinc pulse shaping filter \cite{ArpanDas24}

{\vspace{-4mm}
\small
\begin{eqnarray}
w_{tx}(\tau, \nu) =  \Omega_{\tau} \Omega_{\nu} \sqrt{BT} \text{sinc}(B \tau) \, \text{sinc}(T \nu)  e^{-(\alpha_{\tau} B^2 \tau^2 + \alpha_{\nu} T^2 \nu^2)},
\end{eqnarray}\normalsize}where $\alpha_{\nu} = \alpha_{\tau} =  0.044$ and $\Omega_{\nu} = \Omega_{\tau} = 1.0278$ ensures that $x(t)$ has $99$ percent energy in duration $T$ and bandwidth $B$, and also that the pulse shaping filter has unit energy, i.e., $\iint \vert w_{tx}(\tau, \nu) \vert^2 d\tau d\nu = 1$.

The Zak-OTFS parameters are $M=31, N=37$, $\nu_p = 30$ kHz, $\tau_p = 33.33 \mu s$, $B = M \nu_p =  0.93$ MHz, $T = N \tau_p = 1.23$ ms. The Rician factor is $K=10$. We simulate three different scenarios, (i) when the non-zero tap $(\tau_2, \nu_2)$ lies on the information lattice (i.e., on-grid), (ii) when $(\tau_2, \nu_2)$ is off-grid and, (iii) when $(\tau_2, \nu_2)$ is near a lattice point.
    \begin{figure}
    \vspace{-2mm}
     \centering
\includegraphics[width=7.5cm,height=5.2cm]{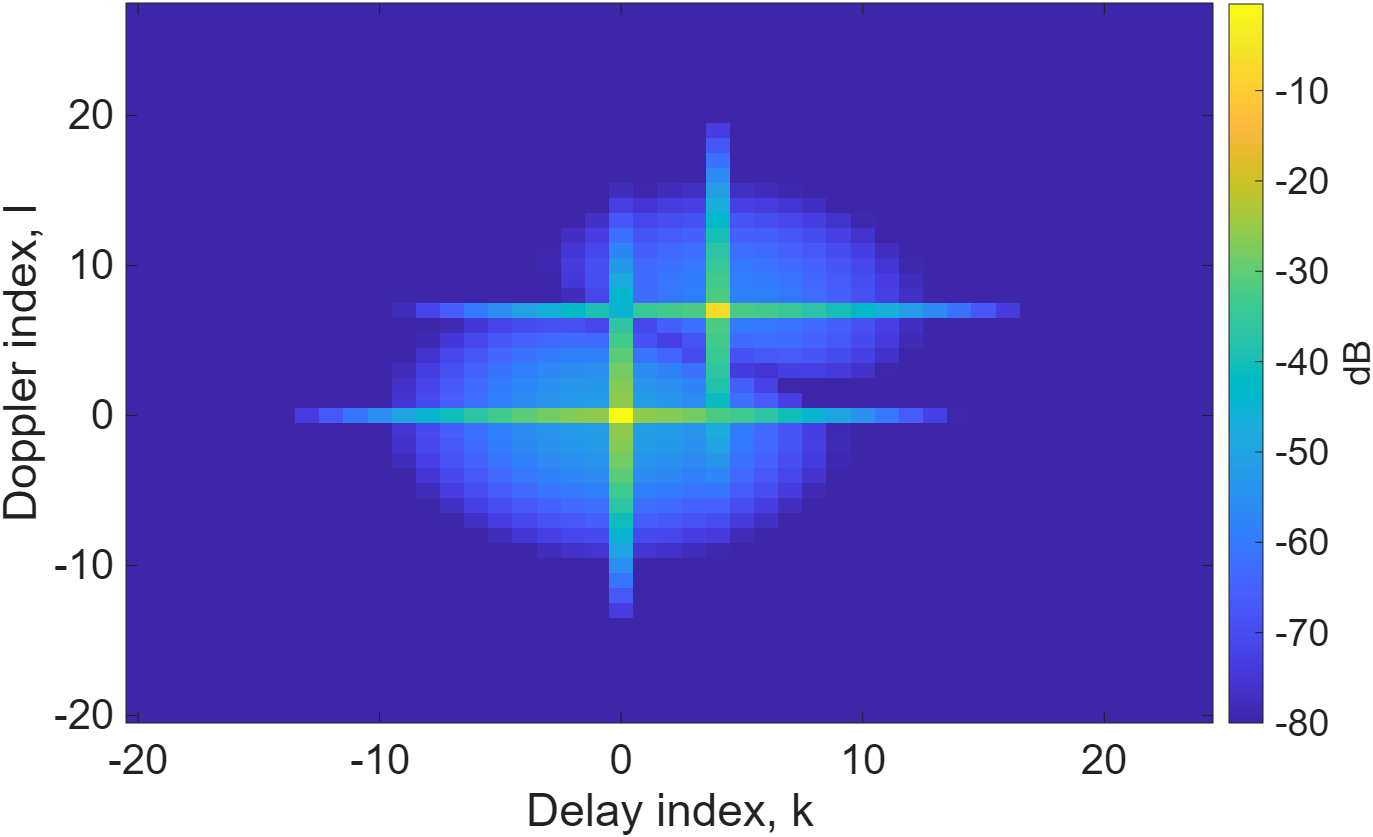}
        \hspace{-6mm}
        \vspace{-2mm}
        \caption{Heatmap of $\vert h[k,l] \vert^2$ for $(\tau_2, \nu_2) = (4/B, 7/T)$ (on-grid).}
        \vspace{-2mm}
        \label{fig1}
    \end{figure}
    
For the first scenario, we consider $\tau_2 = 4/B$ and $\nu_2 = 7/T$. Since this tap lies on-grid, the corresponding $h[k,l]$ has two dominant taps ($N_t = 2$) at $(k,l) = (0,0)$ and $(k,l) = (4,7)$ (see the two bright yellow spots in the heatmap in Fig.~\ref{fig1}). To enable low-complexity single-tap equalization, as discussed in Section \ref{sec3}, we choose the ZC root $u$ such that $(4,7)$ lies on the line $l = -u k (\text{mod} \, MN)$, i.e., $u$ must be such that $(7 + 4u)$ is divisible by $MN = 1147$ which gives $u = 285$ ($285$ and $MN = 1147$ are co-prime).  
    \begin{figure}
     \centering
\includegraphics[width=7.5cm,height=5.2cm]{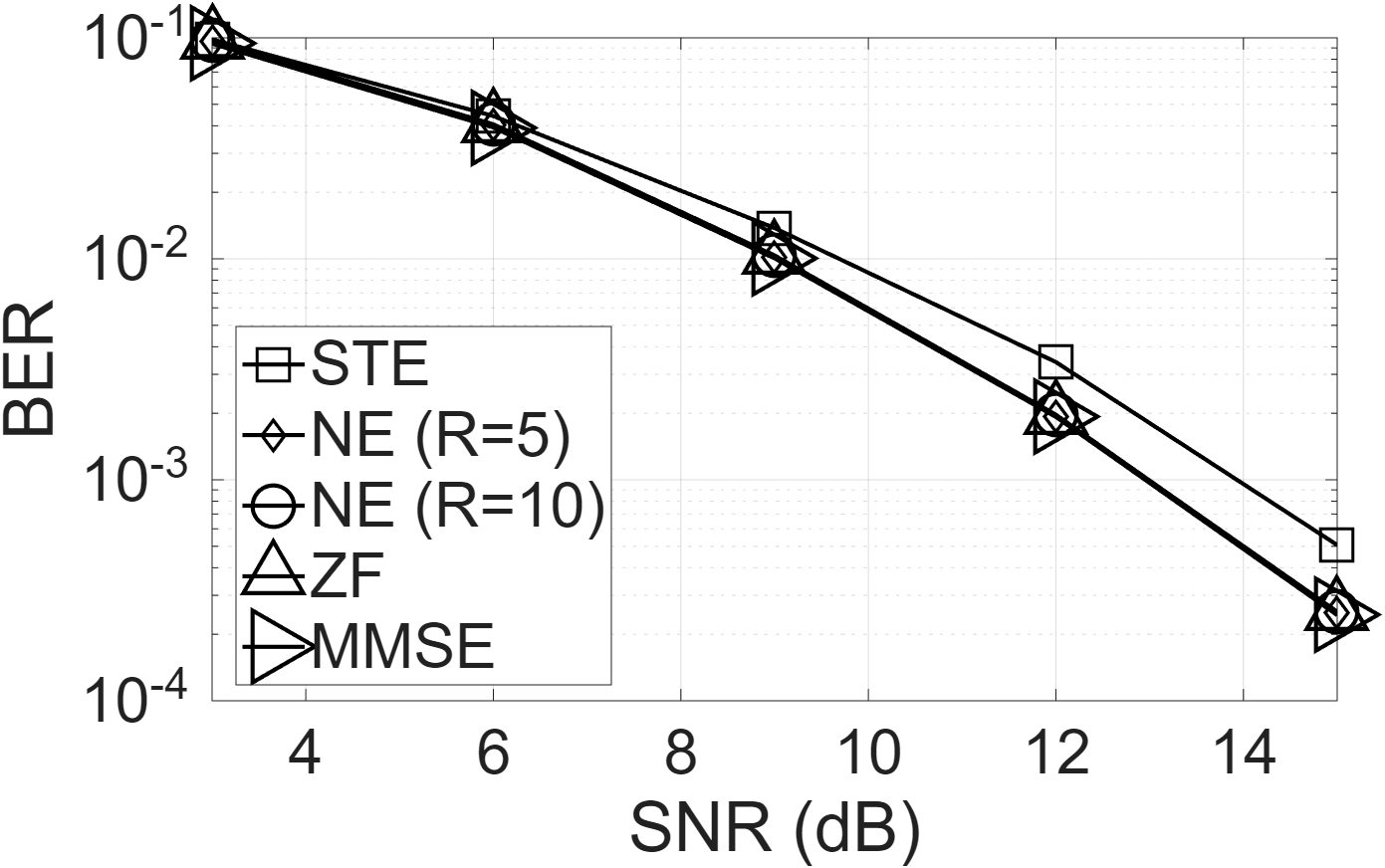}
        \hspace{-6mm}
        \vspace{-3mm}
        \caption{Uncoded $4$-QAM BER vs. SNR (dB). Perfect channel state information. $(\tau_2, \nu_2) = (4/B, 7/T)$ (on-grid), $u=285$.}
        \vspace{-2mm}
        \label{fig2}
    \end{figure}In Fig.~\ref{fig2}, we plot the uncoded $4$-QAM bit error rate (BER) performance of the STE, NE ($R=5$ and $R=10$), the ZF and the minimum mean square error (MMSE) equalizers, as a function of increasing SNR $\gamma$. We consider $u=285$ as it gives waveforms which are eigen-vectors of the channel and therefore there is no inter-carrier interference. Hence, the performance of STE is almost the same as that of the other equalizers. STE BER is slightly inferior, since a small fraction of the channel energy still leaks into DD taps not on the line joining the two dominant taps (see Fig.~\ref{fig1}).

\begin{figure}
    \vspace{-2mm}
     \centering
\includegraphics[width=7.5cm,height=5.2cm]{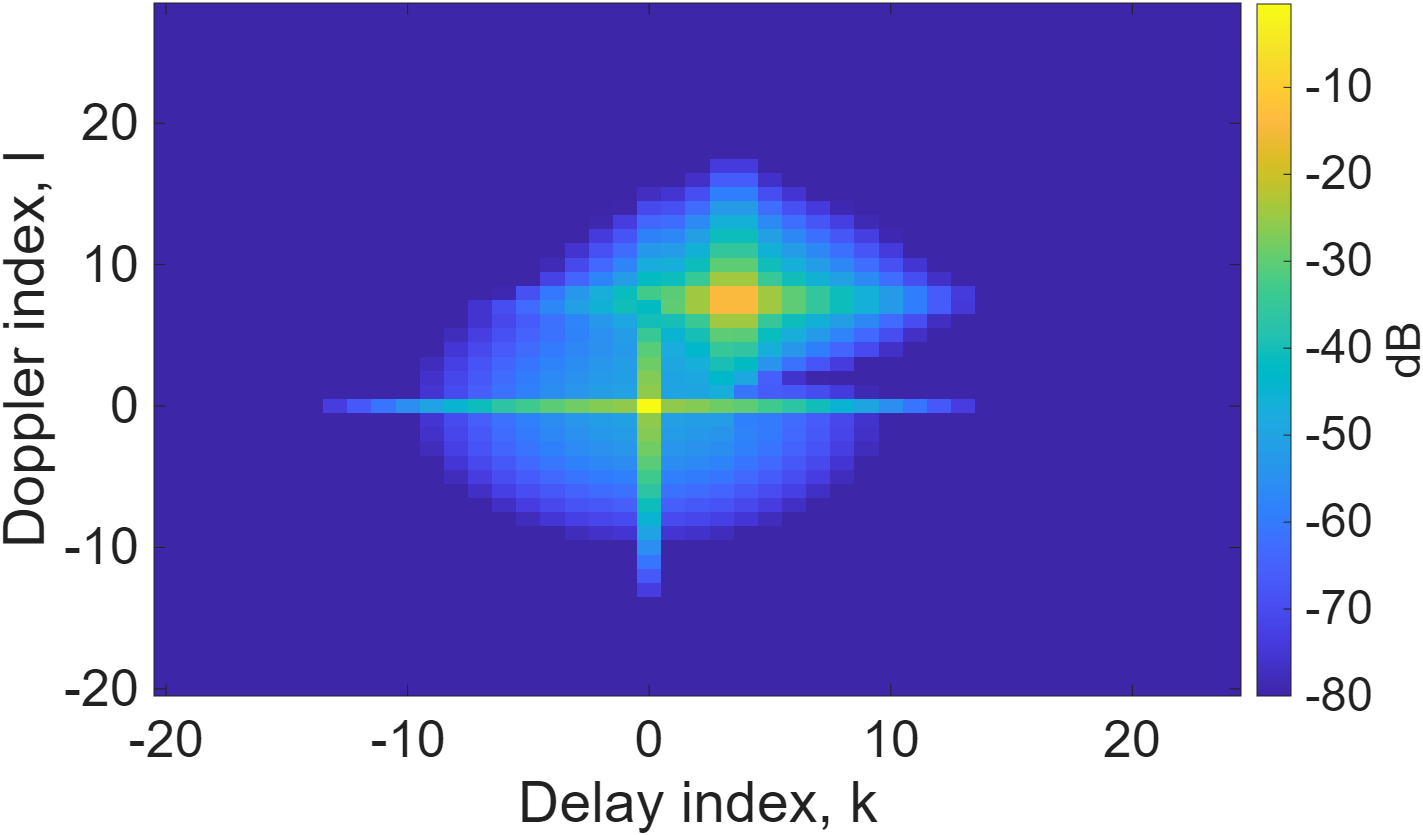}
        \hspace{-6mm}
        \vspace{-1mm}
        \caption{Heatmap of $\vert h[k,l] \vert^2$ for $(\tau_2, \nu_2) = (3.5/B, 7.5/T)$ (off-grid).}
        \vspace{-1mm}
        \label{fig3}
        \vspace{-1mm}
    \end{figure}

    Next, we consider the second scenario where $(\tau_2, \nu_2 ) = (3.5/B, 7.5/T)$, i.e., off-grid tap. The heatmap of $\vert h[k,l] \vert^2$ for a channel realization is shown in Fig.~\ref{fig3}. 
    Since the second path has a delay shift exactly in the middle of the third and the fourth delay tap, and Doppler shift exactly in the middle of the seventh and eighth Doppler tap, its energy is spread equally over four taps (see Fig.~\ref{fig3}) and therefore they cannot lie simultaneously on the same line. However in the continuous DD domain $(3.5/B, 7.5/T)$ lies on the line which passes through $(7/B, 15/T)$. The continuous DD tap $(7/B, 15/T)$ corresponds to the discrete DD tap $(k,l) = (7,15)$ and lies on the line $l = -u k (\text{mod} MN)$ with $u = 981$.   
    \begin{figure}
    \vspace{-2mm}
     \centering
\includegraphics[width=7.5cm,height=5.2cm]{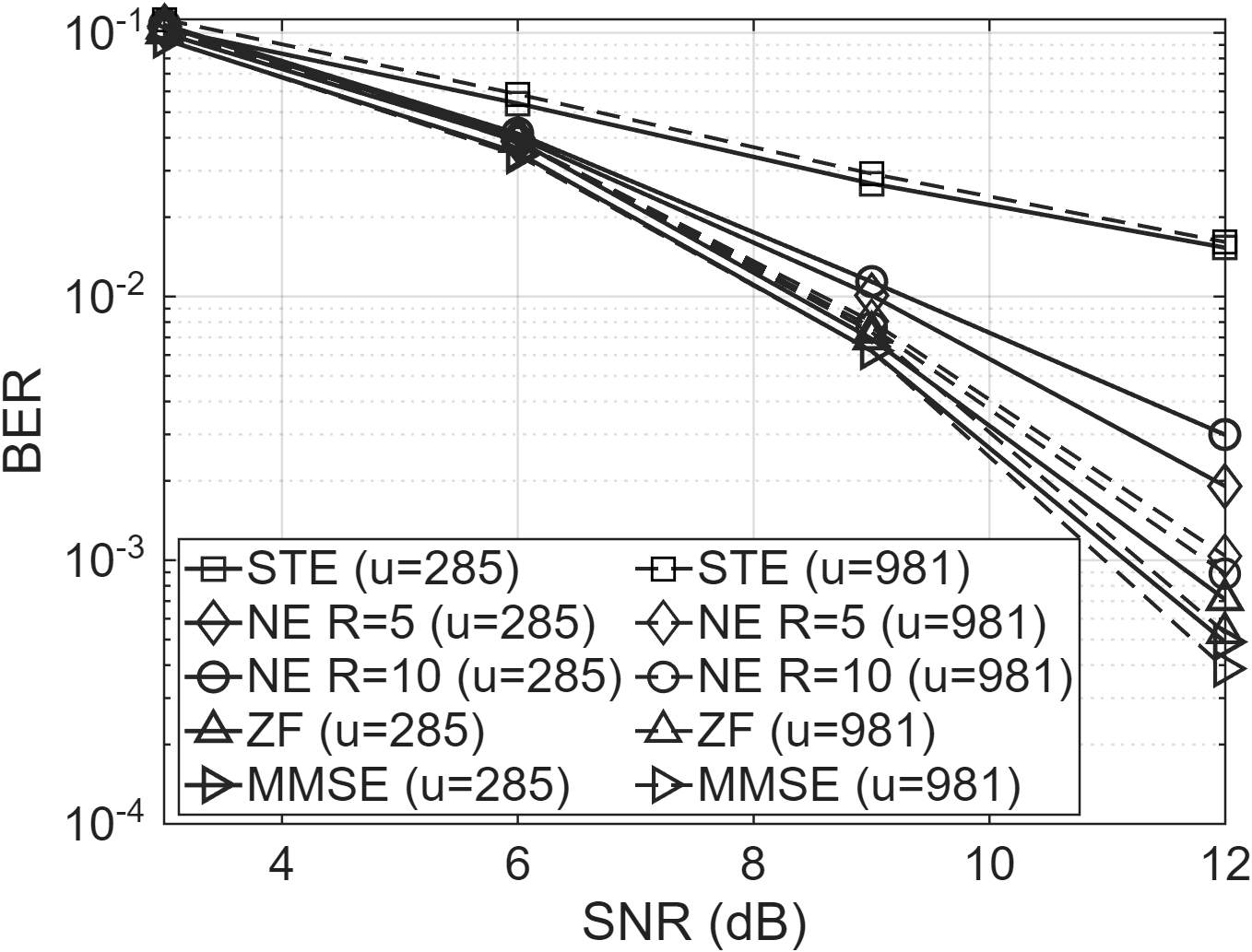}
        \hspace{-6mm}
        \vspace{-1mm}
        \caption{Uncoded $4$-QAM BER vs. SNR (dB). Perfect channel state information. $(\tau_2, \nu_2) = (3.5/B, 7.5/T)$ (off-grid), $u=285$ and $u=981$.}
        \vspace{-1mm}
        \label{fig4}
        \vspace{-1mm}
    \end{figure}

    In Fig.~\ref{fig4} we plot the uncoded $4$-QAM BER vs. SNR for the second scenario with $(\tau_2, \nu_2) = (3.5/B, 7.5/T)$, for both $u=285$ (corresponding to the line passing through $(k,l) = (4,7)$) and $u=981$ (which in the continuous DD domain corresponds to the line passing through $(\tau_2, \nu_2)$). Since the second path is off-grid, it is not possible to choose a line which passes through the five dominant taps (the tap at $(0,0)$ and the other four taps corresponding to off-grid $(\tau_2,\nu_2)$), and therefore the STE performance for both values of $u$ is poor compared to that of the other equalizers.
    The Neumann equalizer although better than STE, has inferior performance when compared to ZF and MMSE. This is because, due to the randomness of path gain $h_2$, ${\bf H}_u$ is not always diagonally dominant and so convergence of NE is not always guaranteed. The performance of NE is better for $u=981$ when compared to that for $u=285$, since in the continuous DD domain $u=981$ corresponds to the line passing exactly through $(\tau_2, \nu_2)$ whereas the line corresponding to $u=285$ does not.
    \begin{figure}
     \centering
\includegraphics[width=7.5cm,height=5.2cm]{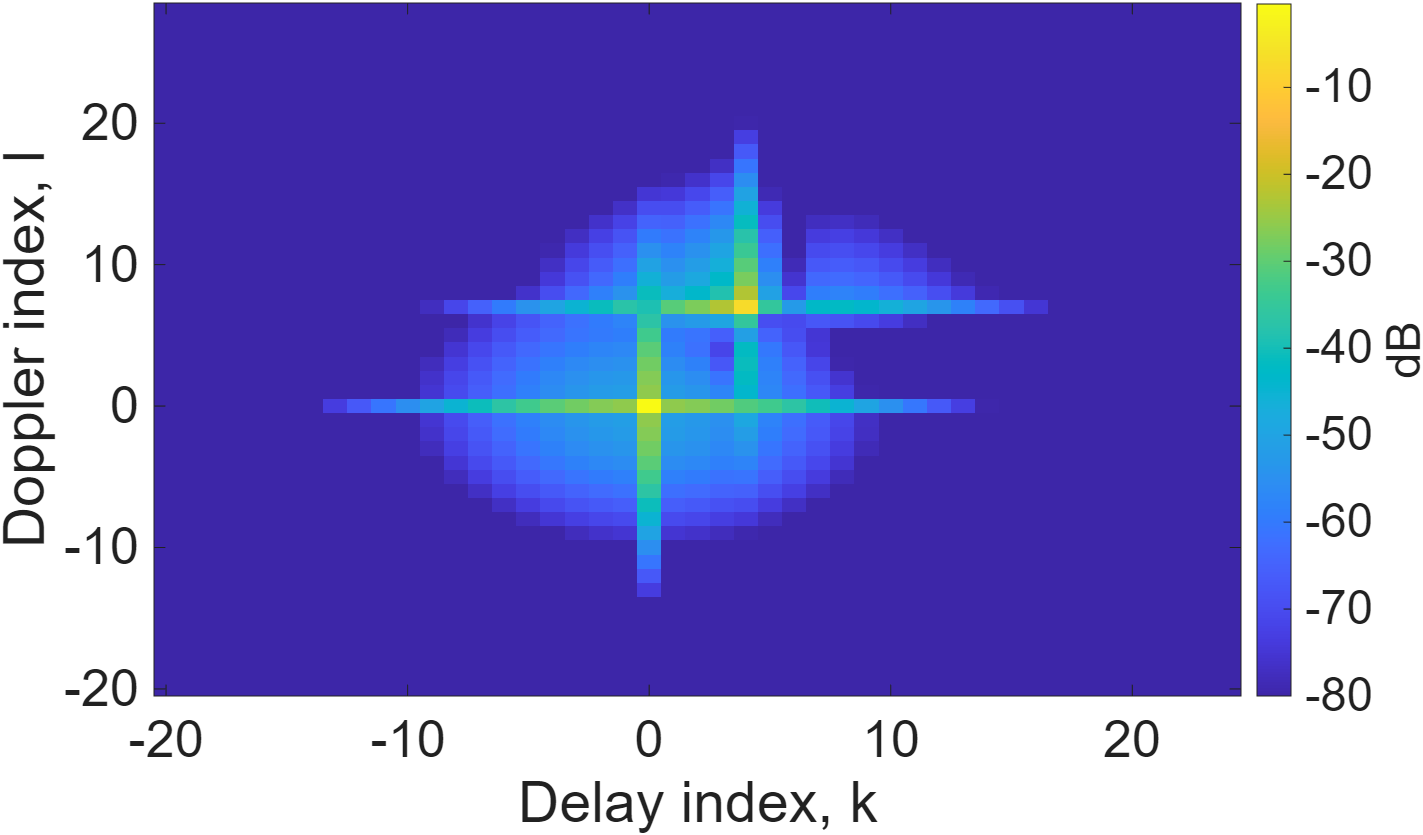}
        \hspace{-6mm}
        \vspace{-1mm}
        \caption{Heatmap of $\vert h[k,l] \vert^2$ for $(\tau_2, \nu_2) = (3.9/B, 7.1/T)$ (near on-grid).}
        \vspace{-1mm}
        \label{fig5}
        \vspace{-1mm}
    \end{figure}
    
    Next, we consider the third scenario where $(\tau_2 , \nu_2) = (3.9/B, 7.1/T)$ which is not exactly on the grid, but very close to it.
    The heatmap of $\vert h[k,l] \vert^2$ is plotted in Fig.~\ref{fig5}. Since the DD shift of the second path is close to $(4/B, 7/T)$, we see a bright spot at $(k,l) =(4,7)$ which also leaks energy to adjacent taps along delay and Doppler since it is not exactly on-grid. Since most of the channel energy is still limited to the two taps at $(0,0)$ and $(4,7)$ we choose the line joining these two taps and choose $u=285$, same $u$ as that for the on-grid channel in the first scenario.
    \begin{figure}
    \vspace{-2mm}
     \centering
\includegraphics[width=7.5cm,height=5.2cm]{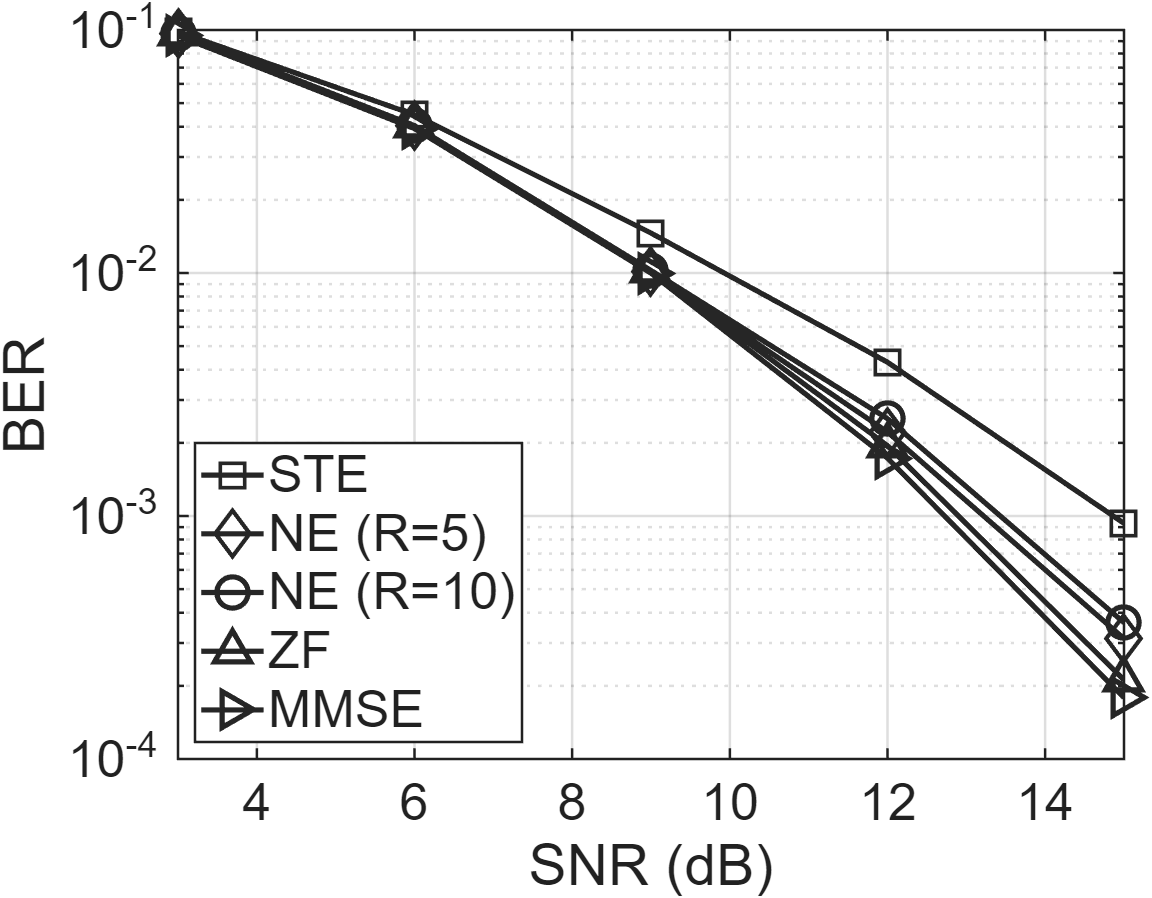}
        \hspace{-6mm}
        \vspace{-1mm}
        \caption{Uncoded $4$-QAM BER vs. SNR (dB). Perfect channel state information. $(\tau_2, \nu_2) = (3.9/B, 7.1/T)$ (near on-grid), $u=285$.}
        \vspace{-1mm}
        \label{fig6}
        \vspace{-1mm}
    \end{figure}

In Fig.~\ref{fig6} , we plot the uncoded $4$-QAM BER vs. SNR for the third scenario where $(\tau_2, \nu_2) = (3.9/B, 7.1/T)$ (near on-grid). Since most of the channel energy in $h[k,l]$ is still dominated by the taps at $(0,0)$ and $(4,7)$, the ZC carrier waveforms with $u=285$ are almost eigen-vectors of the channel and therefore inter-carrier interference is smaller than that for the second scenario. This is why the performance gap between STE and ZF/MMSE is smaller than that in the second off-grid scenario (see Fig.~\ref{fig4}). The performance of NE is close to ZF/MMSE performance since the probability of ${\bf H}_u$ not being diagonally dominant is much lesser than that in the second scenario ($u=285$). Note that there is still a finite probability that ${\bf H}_u$ is not diagonally dominant (due to randomness in the gain $h_2$ of the second path), and therefore NE does not converge for a small fraction of channel realizations.
    \begin{figure}
     \centering
\includegraphics[width=7.5cm,height=5.2cm]{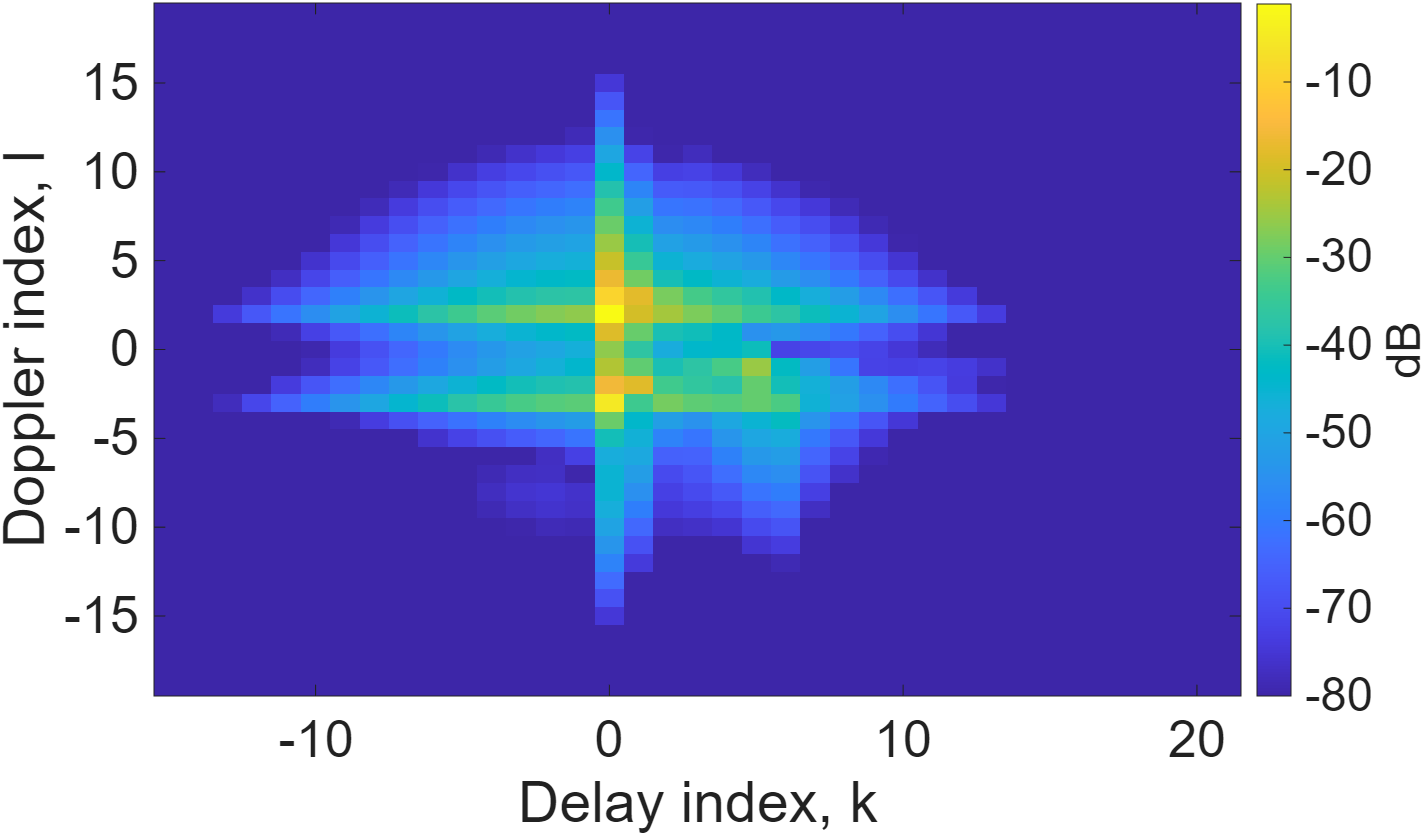}
        \hspace{-6mm}
        \vspace{-1mm}
        \caption{Heatmap of $\vert h[k,l] \vert^2$ for the 3GPP TDL-D channel model. RMS delay spread of $0.3 \mu s$ and $\nu_{max} = 5.55$ kHz.}
        \vspace{-1mm}
        \label{fig7}
        \vspace{-1mm}
    \end{figure}
    
We also consider the 3GPP standardized TDL-D LOS channel model \cite{3gpptr38901} with root mean square (RMS) delay spread of $0.3 \mu s$ and $\nu_{max} = 5.55$ kHz. The zero delay path is Rician distributed, i.e., it has a deterministic component with fixed gain and fixed Doppler shift of $0.7 \nu_{max}$ and the random component has Rayleigh distributed gain with random Doppler shift $\nu_{max} \cos(\theta)$ ($\theta$ is uniformly distributed in $[0 \,,\, 2 \pi)$). For the zero delay path, the K factor is $13.3$ dB. The heatmap of $\vert h[k,l] \vert^2$ for a realization of the TDL-D channel is shown in Fig.~\ref{fig7}. Indeed, most of the channel energy is localized to taps on the Doppler axis (i.e., zero delay line).
There is no $u$ for which $l = -u k (\text{mod} \, MN)$ coincides with the delay zero line $k=0$.
Since a large fraction of the channel energy is localized to the zero delay taps, we choose waveforms which are eigen vectors of a zero delay channel. These waveforms are simply the $MN$ impulses in discrete-time, i.e., $s_q$ is carried by the impulse at $n= q$ which is simply the periodic signal $\sum\limits_{i \in {\mathbb Z}} \delta[n - q - i MN]$, $q=0,1,\cdots, MN-1$. In our notation this corresponds to ${\bf Z}_u = {\bf I}$. 
    \begin{figure}
    \vspace{-2mm}
     \centering
\includegraphics[width=7.5cm,height=5.2cm]{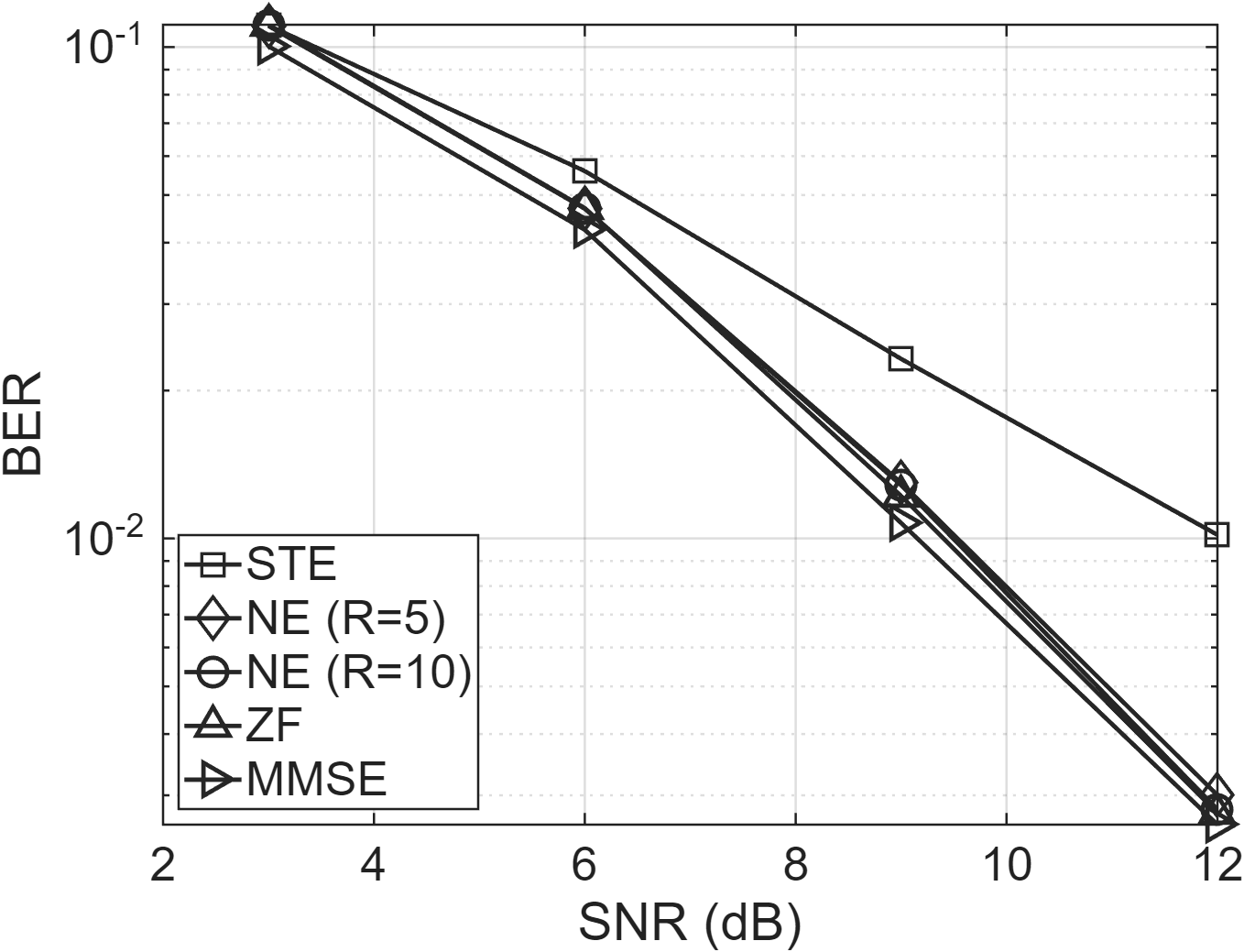}
        \hspace{-6mm}
        \vspace{-1mm}
        \caption{Uncoded $4$-QAM BER vs. SNR (dB). Perfect channel state information.  3GPP TDL-D channel. RMS delay spread $0.3 \mu s$, $\nu_{max} = 5.55$ kHz. ${\bf Z}_u = {\bf I}$.}
        \vspace{-1mm}
        \label{fig8}
        \vspace{-1mm}
    \end{figure}
    
In Fig.~\ref{fig8} we plot the uncoded $4$-QAM BER vs SNR performance for the TDL-D channel model described above, with ${\bf Z}_u = {\bf I}$. While the columns of ${\bf Z}_u = {\bf I}$ are eigen-vectors for the channel taps on the zero delay line (i.e., Doppler axis), they are not eigen for the small fraction of the channel energy on taps which are not on the zero delay line (see heatmap in Fig.~\ref{fig7}). STE is unable to equalize the inter-carrier interference resulting from the channel energy in taps which are not on the zero delay line, and therefore it has inferior BER performance compared to the other equalizers. NE has similar performance as ZF/MMSE since ${\bf H}_u$ is mostly diagonally dominant.

\section{Conclusion}
\label{sec6}
The paper presented a low-complexity equalization framework for Zak-OTFS that selects an orthonormal carrier-waveform basis matched to the delay-Doppler geometry of the channel. In particular, when all channel delay-Doppler taps lie on the line $l=-uk \pmod{MN}$, the ZC carrier waveforms in the basis with root $u$, obtained from the Zak-OTFS pulsone basis by applying the unitary GDAFT, are eigenvectors of the effective channel matrix. Hence, the effective channel is diagonal in this ZC basis and single-tap equalization is enabled. More generally, when the effective channel is diagonally dominant in the selected basis, the proposed Neumann equalizer mitigates the interference caused by its off-diagonal entries. Simulation results for the two-path and 3GPP TDL-D LOS channel models demonstrate the performance of the proposed basis selection and Neumann equalization framework for on-grid and fractional delay-Doppler shifts. The results show that the Neumann equalizer improves upon single-tap equalization and, in the scenarios where the effective channel is mostly diagonally dominant, has performance similar to the ZF and MMSE equalizers.

\end{document}